# ON THE TIDAL ENVIRONMENT OF AN OUTWARDLY MIGRATING F-RING

Phil J. Sutton

psutton@lincoln.ac.uk

University of Lincoln, School of Mathematics and Physics, Brayford Pool, Lincoln, LN6 7TS, UK

Corresponding author: Phil Sutton, psutton@lincoln.ac.uk

# Abstract

Saturn's F-ring is a unique, narrow ring that lies (radially) close to the tidally disruptive Roche limit of water ice for Saturn. Significant work has been done that shows it to be one of the most dynamic places in the Solar System. Aggregates that are fortunate enough to form constantly battle against the strong tidal forces of Saturn and the nearby moons Prometheus and Pandora, which act to gravitationally stir up ring material. Planetary rings are also known to radially spread. Therefore, as the F ring lies at the edge of the main rings, we investigate the effect of an outwardly migrated F ring and its interaction with Prometheus. An increase in the maximum number density of particles at the channel edges is observed with decreasing local tidal environment. Radial velocity dispersions are also observed to fall below the typical escape velocity of a $150m$ icy moonlet ($< 10\ cm\ s^{-1}$) where density is enhanced, and are gravitationally unstable with Toomre parameters $Q < 2$. Additionally, in locations of the ring where $Q < 2$ is observed, more particles are seen to fall below or close to the critical Toomre parameter as the radial location of the ring increases.

1. INTRODUCTION

There have been substantial advancements in our understanding of the dynamics of planetary ring systems (Hyodo & Ohtsuki 2014; Attree et al 2014; Rein & Latter 2013; Hedman et al 2013; Attree et al 2012; Hedman et al 2011; Beurle et al 2010; Rein & Papaloizou 2010; Tiscareno et al 2010; Lewis & Stewart 2009; Weiss et al 2009; Murray et al 2008; Thomson et al 2007; Hedman et al 2007; Murray et al 2005; Salo et al 2001). A curious and interesting ring is Saturn's F ring, a narrow ring located at the edge of the main rings, which is seemingly shepherded by two nearby moons Prometheus and Pandora. Situated at the edge of the main rings, the F ring is at a unique location near the Roche limit for water ice around Saturn. The Roche limit is described as the minimum distance an object can be from the primary mass before it is tidally torn apart. For a purely fluid satellite consisting of ice around a planet this is given as (Chandrasekhar 1969),

$$d = R \cdot 2.46 \sqrt[3]{\left(\frac{\rho_M}{\rho_m}\right)} \qquad [1]$$

where $R$ is the radius of the primary mass (in our case the central planet), $\rho_M$ is the density of the planet and $\rho_m$ is the density of the satellite. Assuming a value for the density of water ice of $0.934\ gcm^{-3}$ in line with previous work (Dubinski 2017; Salo 1992), for Saturn this approximates to $d = 135,000 km$, which is very close to the location of the F ring.

Thus, the F ring is a very dynamic system for any aggregates which do manage to form. These aggregates quickly find themselves fighting against the strong tides from Saturn, and interact strongly with nearby moons (Cuzzi et al 2014). A rather puzzling fact is that the moon Pan (Horn et al 1996) and hundreds of propeller structures caused by small $\sim 100m$ sized moonlets (Seiß et al 2017; Lewis & Stewart 2009; Tiscareno et al. 2008; Sremčević et al 2007), exist within the Roche limit for Saturn. It is unlikely that these moons and moonlets formed in their current locations within the rings; it is more likely that other processes must be responsible for their existence.

Motivation for this work stems from the fact that planetary rings are known to radially spread through a process known as viscous spreading (Charnoz et al 2011; Charnoz et al 2010). A change in radial position of a ring naturally alters the local tidal environment. The F ring lies at a unique location close to the Roche limit around Saturn, but also at the edge of the main rings. Therefore, its location is of interest to the radial spreading case of planetary rings. Currently it is unclear if the F ring and its shepherd moons formed at this radial location, because it is close to the Roche limit or purely a coincidence. Hyodo & Ohtsuki (2015) showed that the F ring and its shepherd moons were a natural outcome of a collision between two satellites at its current location. If such a collision occurred at a larger radial distance than the current location of the F ring, it would place it in a lower tidal environment but still with nearby shepherd moons. Thus, were a ring to exist further beyond the Roche limit of the host planet, the tides would not be strong enough to compete against the formation of aggregates. Narrow rings and arcs have been found to exist around Saturn for example the G ring (Hedman et al 2007), although these are likely to be dynamically short lived. Therefore, narrow shepherded moons could exist at larger radial distances from the host planet. If nearby moons disrupt these rings, we would naturally assume the rings to be more gravitationally distorted, as they exist at locations where the tides are lower. The decreased tides from the host planet would be beneficial in the formation of aggregates, however, the competing interactions with the shepherd moons might also be more destructive for such aggregates.

## 2.    NUMERICAL METHOD

In this work, we compare three separate tidal environments, including the original F-ring – Prometheus system. The tidal environment of a ring can be changed by altering one of two parameters, the first being to reduce the mass of the central planet, such that the gravitational tides are reduced at the same distance, and the second being to change the radial location of the ring itself, so that the gravitational tides are less severe. Since it is known that planetary rings like those around Saturn periodically spread outwards to accrete moons when beyond the Roche Limit (Charnoz et al 2011; Charnoz et al 2010), it is then natural to assume the second scenario for our work. The initial starting model we employed is the same as previously investigated (Sutton & Kusmartsev 2013, 2014 & 2016).

Saturn is placed at the origin of our system of coordinates, with a zero-magnitude velocity vector. The equations for initial positions of all particles ($\boldsymbol{R_s} = Saturn$, $\boldsymbol{R_p} = Prometheus$, $\boldsymbol{R_j} = ring\ particles$) are taken from Murray et al 2005 and can be shown as:

$$\boldsymbol{R_s} = [0,\ 0,\ 0] \tag{2}$$

$$\boldsymbol{R_p} = [139{,}671\ km,\ 0,\ 0] \tag{3}$$

$$\boldsymbol{R_j} = [r \cdot \cos\theta,\ r \cdot \sin\theta, 0] \tag{4}$$

where $r$ represents the radial position of ring particles from Saturn, and $\theta$ the angular position (between $-\pi/4 - \pi/4$) of ring particles around Saturn. We used a ring patch to only model part of the ring in order to reduce computational costs, thus only a $\pi/2$ section of the F ring is modelled.

The radial positions of ring particles are distributed over four groups which represent the background sheet of particles', the inner strand, the central core, and the outer strand. The inner ring boundary and width is then given as:

$$r_1 = 139{,}876\ km \qquad w_1 = 700\ km$$
$$r_2 = 140{,}049\ km \qquad w_2 = 70\ km$$
$$r_3 = 140{,}214\ km \qquad w_3 = 20\ km$$
$$r_4 = 140{,}299\ km \qquad w_4 = 30\ km$$

Therefore, outer edge ring particles in each group are radially located at $r = r_n + w_n$. Particle velocities are then derived assuming Keplerian motion, with Prometheus placed at the periapsis of its orbit for $T = 0$, as taken from Spitale et al (2006).

The fundamental code utilised for the integration was GADGET2 (Springel 2005), which is a smoothed particle hydrodynamic (SPH) code. As a result, ring particles have smoothing lengths assigned just above their physical size according to an internal density of ice $0.934\ gcm^{-3}$. We use the density of water ice because Cassini has confirmed that Saturn's rings are predominantly composed of water ice (Hedman et al 2013; Esposito 2010), therefore assuming this density in our models seems reasonable. Additionally, it should be noted that the density of nearby icy moons Pan, Atlas, Daphnis, Pandora, and Prometheus is known to be $< 0.5 \, gcm^{-3}$ (Thomas 2010), which is less than water ice. Therefore, an element of uncertainty surrounds the density of ring particles used here, even when the material composing the rings and moons is well known. Individual particles in our models are separated by $\sim 1 km$, and increasing smoothing lengths by assuming $\rho_{ring} < 0.5 \, gcm^{-3}$ still results smoothing lengths orders of magnitude less than the distance separating particles.

Gadget-2 assumes collision-less dynamics instead of treating particles with physical collisions, thereby reducing gravitational forces within a set smoothing length using a smoothing kernel. The particle types employed within Gadget-2 only considered gravitational forces, and were not subjected to any additional hydrodynamical forces, as is common for cosmological simulations with Gadget-2.

The F ring is quite a diffuse ring with little mass, aside from the moonlet belt in the core (Cuzzi et al 2014). Since the surface density of a ring is the mass of the ring divided by the area, we calculate the surface density for each ring, strand, and core in our models as:

| | |
|---|---|
| Background sheet of particles | $\sim 1 \cdot 10^{-9} \, kg \, m^{-2}$ |
| Inner strand | $\sim 1.5 \cdot 10^{-8} \, kg \, m^{-2}$ |
| Central core | $\sim 5 \cdot 10^{-8} \, kg \, m^{-2}$ |
| Outer strand | $\sim 3 \cdot 10^{-8} \, kg \, m^{-2}$ |

These values are very low compared with the values in the main rings of orders $\sim 100 - 400 \, kg \, m^{-2}$. Even though our models differ from Beurle et al 2010, as we include self-gravity between ring particles, the very low surface densities used are not expected to significantly alter the dynamics over the short time periods we consider. Assuming the surface densities used in our models, approximate timescales for 100m sized moonlets to form would be orders of magnitudes greater than those calculated later in the discussion section ($T_{grow} \sim 0.25, 0.27, 0.31 \, yr$). Therefore, we do not believe that the inclusion of self-gravity in our models significantly alters to the dynamical outcome when comparing with Beurle et al 2010.

It should also be noted that since we increase the semi-major axis of the F ring, the total area that the ring covers also increases. Thus, we ensure that the surface densities and particle number densities within in each part of the ring remain the same in each of the models and are comparable with one another. This is done by increasing the total number of particles in each angular segment of the ring modelled so that number and surface densities are constant.

Additionally, the orbital periods of particles also change as the semi-major axis is increased. Therefore, instead of using an absolute time when discussing the models, we normalise to the orbital period of Prometheus. This also ensures that the features discussed occur at the same configurations, even though the actual interactions will be slower (since the orbital period can be given as $T = 2\pi\sqrt{GM}$. Radial velocity dispersions or ring particles are calculated by measuring the deviation of individual particles from the RMS (Root Mean Squared) of the radial velocities of nearby particles. Two-dimensional visualisations of radial velocity dispersions, Toomre Parameters, and densities are generated with SPLASH (Price 2007). We do this to understand where maximum and minimum dispersions occur, and are just as important as the magnitudes of the dispersions measured.

## 3. NUMERICAL MODEL RESULTS

Beurle et al (2010) investigated the streamer-channel formations created by the close passage of Prometheus to the F ring. It was found that the maximum number density of particles in their model increased to a maximum when the channel formations were at their most open. This corresponds to when Prometheus is at its closest to the F ring, or at the apocentre of its orbit. The channel edges were found to be the locations of the highest densities, and matched the locations of moonlets in the streamer-channel formations as observed by Cassini, thus making a striking connection between perturbing moon and newly formed coherent objects in the F ring. We performed the same analysis of the maximum number density over the first 5 orbital periods for our three models, shown in Fig 1. What is striking is that by taking only the maximum number density into account, we find that the maximum number densities at the channel edges increase with increasing radial location of the F ring from Saturn (decreasing tidal environment). Three orbital periods after the initial encounter of Prometheus, the maximum number densities measured were $21, 23 \& 28$ (from an initial value of 18) for the models located at $140,000 km, 150,000 km \& 160,000 km$ respectively. Figures 2, 3, & 4 show snapshots of each model with density spatially visualised in a surface rendered plot at a time $T = 3.5$. Although not obvious from each figure, the areas disrupted by Prometheus increase with increasing distance from Saturn. The degree of the disruption in the ring extends further and the channels are slightly wider.

In agreement with Beurle et al 2010, the maximum number densities are observed at the channel edges in all models. This is seen to increase with increasing radial distance of the F ring from Saturn, or with decreasing tidal environment, because of the particles' trajectories overlapping more, leading to more bunching up on the channel edges. In addition to the cycle in maximum number density reported by Beurle et al. 2010, we also note that the models located at $150,000 km \& 160,000 km$ show an overlapping of the core and outer strand. This drives a further increase in particle number densities much earlier than the reported surge after 5 orbital periods and would not have been observed by Beurle et al. 2010.

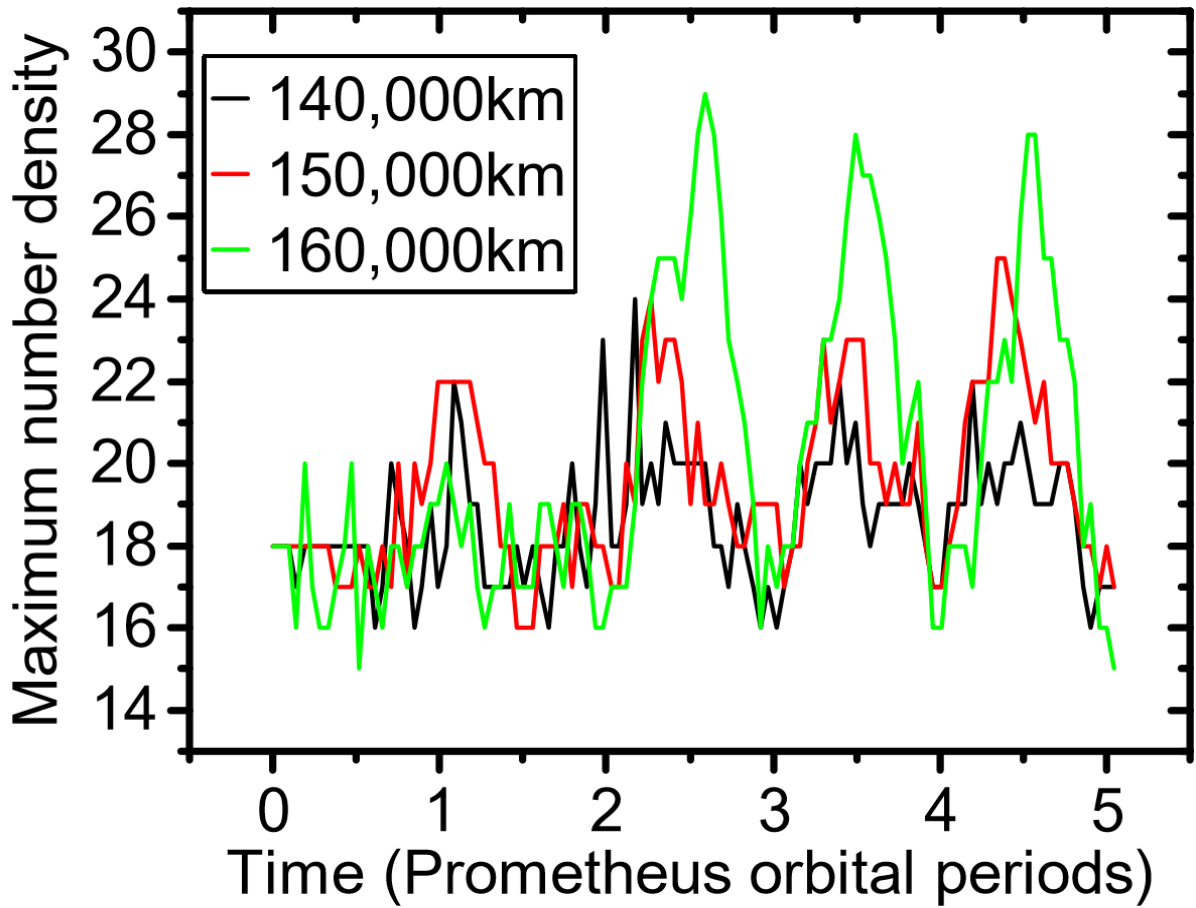

**Figure 1 |** The maximum number density of particles for all three models is plotted with respect to time. The starting maximum number density for each model was set to 18, following Buerle et al 2010. The previously reported enhancement of density occurs when the channels are at their most open, starting most notably at $T = 2.5, 3.5, 4.5$, where $T$ is in Prometheus orbital periods. Density enhancements become more pronounced as the F ring moves radially outwards and resides in an environment where the tides are reduced.

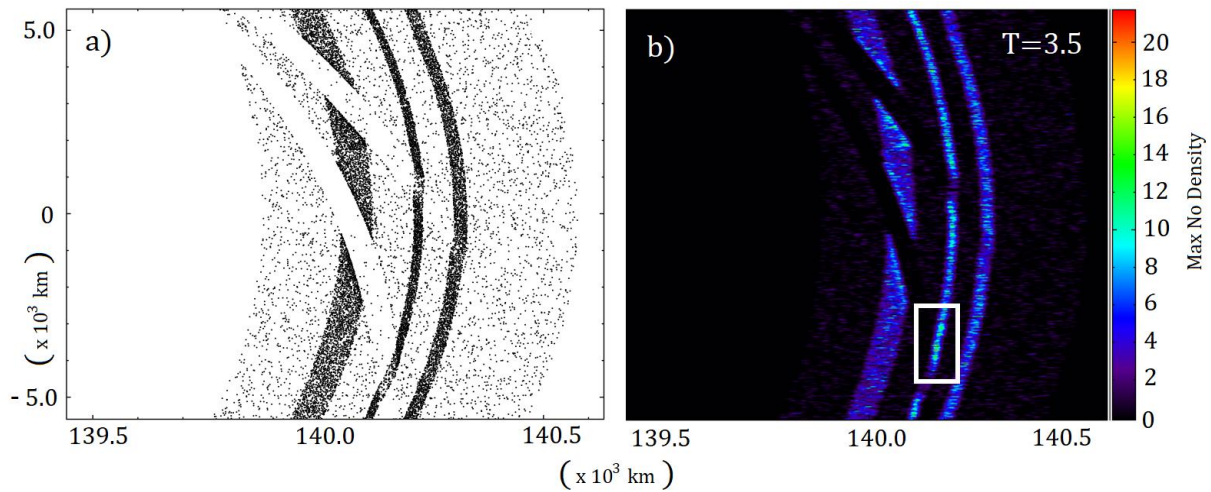

**Figure 2** | Taken at a time $T = 3.5$ since the start of the simulation, where $T$ is in Prometheus orbital periods. a) represents the particle positions for the model taken at the current known location of the F ring (~$140,000 km$), and clearly shows the channel formations at their most open phase. Here Prometheus is out of view and is located above the channels. b) shows the density rendered for the same model, with the area identified with the highest density marked with a white box (Max No. Density = 21).

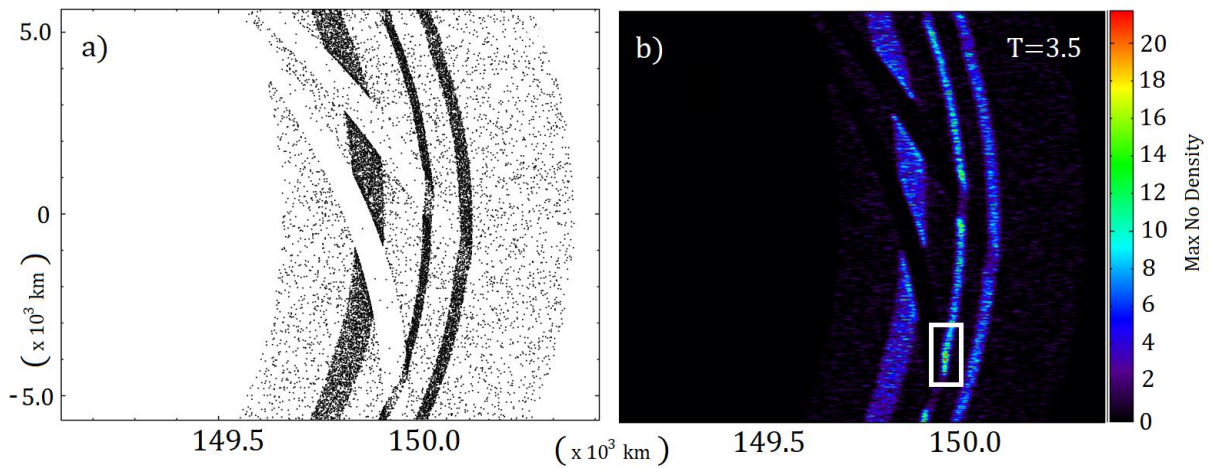

**Figure 3** | Taken at a time $T = 3.5$ since the start of the simulation, where $T$ is in Prometheus orbital periods. a) represents the particle positions for the model taken at location of the F ring that puts the central core at a semi-major axis of $150,000 km$. b) shows the density rendered for the same model, with the area identified with the highest density marked with a white box (Max No. Density = 23).

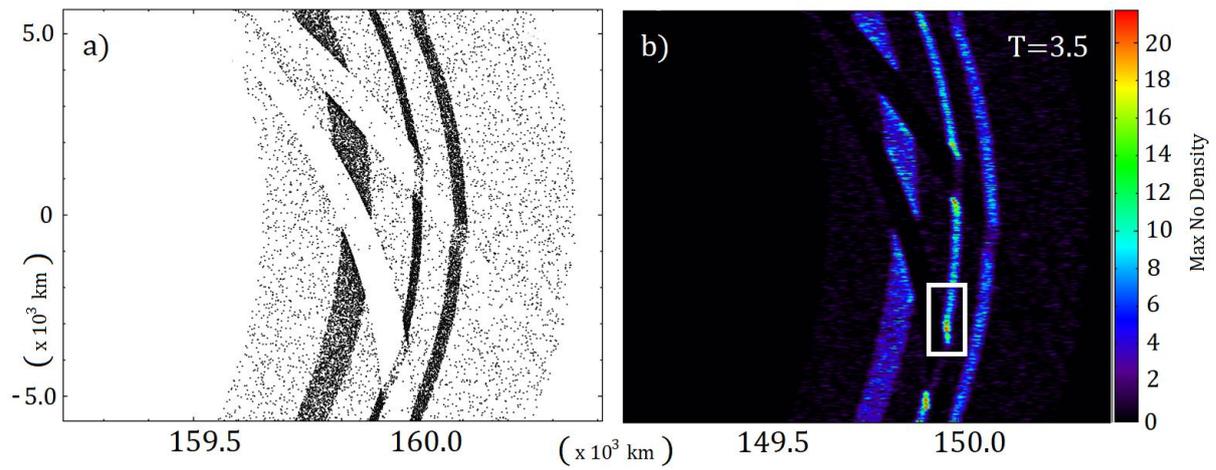

**Figure 4 |** Taken at a time $T = 3.5$ since the start of the simulation, where $T$ is in Prometheus orbital periods. a) represents the particle positions for the model taken at location of the F ring that puts the central core at a semi-major axis of $160,000 km$. b) shows the density rendered for the same model, with the area identified with the highest density marked with a white box (Max No. Density = 28).

## 4.  Local Gravitational Instabilities

We know that Prometheus can produce local enhancements in the F ring's density (Beurle et al. 2010); this is more pronounced in our outwardly migrated models. In conjunction with periodic fluctuations of maximum number densities, radial velocity dispersions played an important role in the further gravitational collapse of these dense regions (Beurle et al 2010). A differentially rotating disk will experience gravitational instabilities when the Toomre parameter $Q \leq 1$ (Toomre 1964).

$$Q = \frac{v_r \Omega}{\pi G \Sigma} \qquad [5]$$

Where $v_r$ is the radial velocity dispersion, $\Omega$ is the orbital angular frequency, $G$ is the gravitational constant, and $\Sigma$ is the surface density of the disk. The Toomre stability parameter ($Q$) evaluates the self-gravity of an astrophysical disk to the shear stresses. In the case of Saturn's rings, the shear stress is in the form of a Keplerian shear with respect to radial position. For the purposes of our work we take $\Sigma$ to be $\sim 50\ kgm^{-2}$ for Saturn's F ring, which is very conservative given the diffuse nature of the F ring. This value of $\Sigma$ is used to calculate the $Q$ values of particles from radial velocity dispersions measured in our models. It does not relate to surface densities in our models which are typically much lower.

When radial velocity dispersions fall below a critical value, the self-gravity of the disk will be greater than the velocities of local encounters between particles. Thus, particles will encounter one another below their mutual escape velocities and form local enhancements in density. In the case of Prometheus and the F ring, enhanced density at the channel edges, coupled with local radial velocity dispersions consistent with $Q \leq 1$, would drive the growth of clumps and moonlets. In the case of the three models we ran, and assuming a surface density of $\Sigma = 50\ kgm^{-2}$, the critical radial velocity dispersions are $8.9 \times 10^{-5}\ ms^{-1}\ (140{,}000\ km),\ 9.9 \times 10^{-5}\ ms^{-1}\ (150{,}000\ km)\ \&\ 1.1 \times 10^{-4}\ ms^{-1}\ (160{,}000\ km)$.

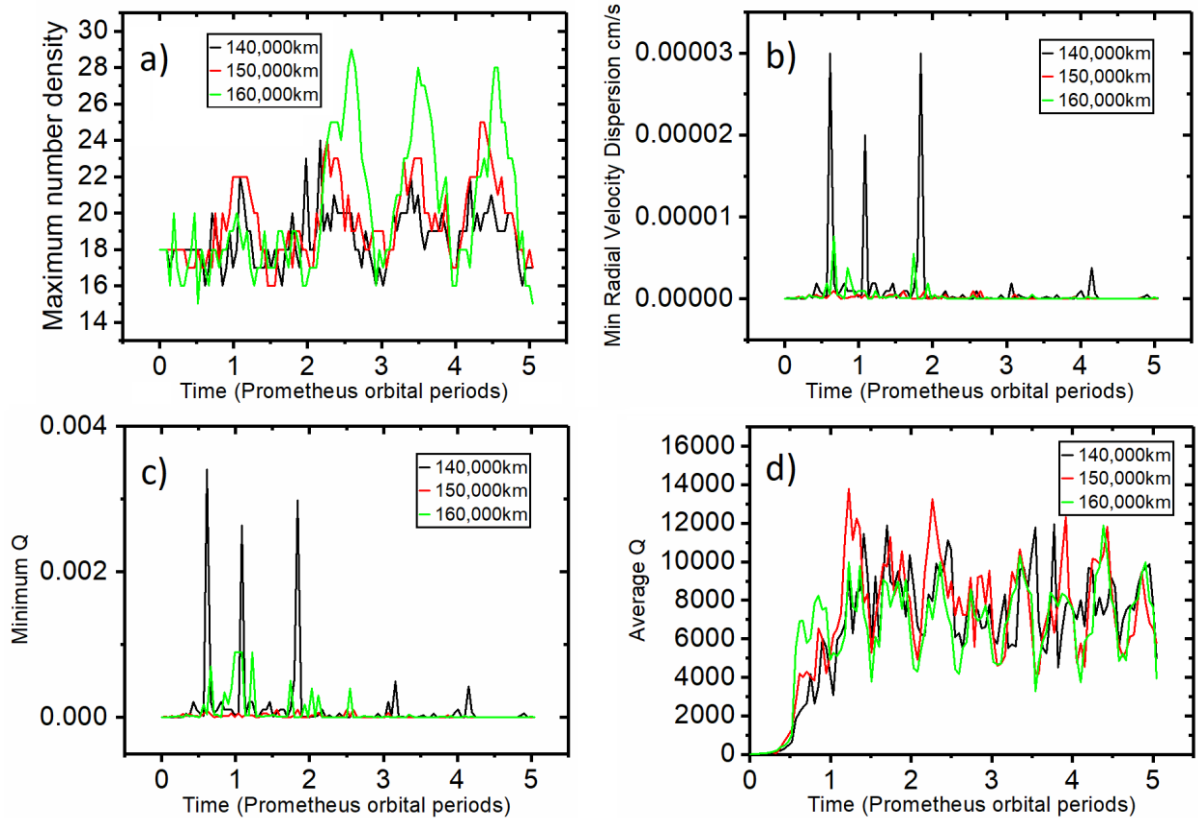

**Figure 5 |** For a ring segment disrupted by Prometheus the a) maximum number density, b) minimum radial velocity dispersion, c) minimum calculated Q value for an F ring with surface density $\Sigma = 50\ kg\ m^{-2}$, d) average Q value for the sampled ring segment is plotted for the first 5 orbital periods of Prometheus after the start of all three models.

Tracking the minimum radial velocity dispersions and minimum Q values is not enough to provide anything conclusive between the three models. The minimum values consistently lay below the critical values for the duration of time monitored. However, as we take the minimum values from a ring segments that are disrupted by Prometheus, there are areas in these segments that show little disruption. Typically, these areas are not near the streamer-channel formations (see the lower left of the F ring in Figures 6, 7 & 8). Average Q values provide more useful information about the environment post encounter. Although average values in the same ring segment are above critical values by orders of magnitude there are hints of a cyclic pattern. In our models, there is evidence to suggest average Q values show two minima per orbital period of Prometheus (Fig 5d), compared with one minima per Prometheus orbital period for maximum number densities (Fig 5a). The average Q values for the disrupted rings do indeed show that radial velocity dispersions tend towards minimum values, while maximum number densities peak for all three models, consistent with Beurle et al 2010. The reason for this is related to the radial movement of an individual particle on an elliptical orbit; there are two locations on an elliptical orbit where the radial velocity of a particle would tend to zero (apocentre and pericentre). More collectively a ring segment disrupted by Prometheus synchronises to the radial movement of Prometheus, which is why we see the cyclic pattern of streamer-channels.

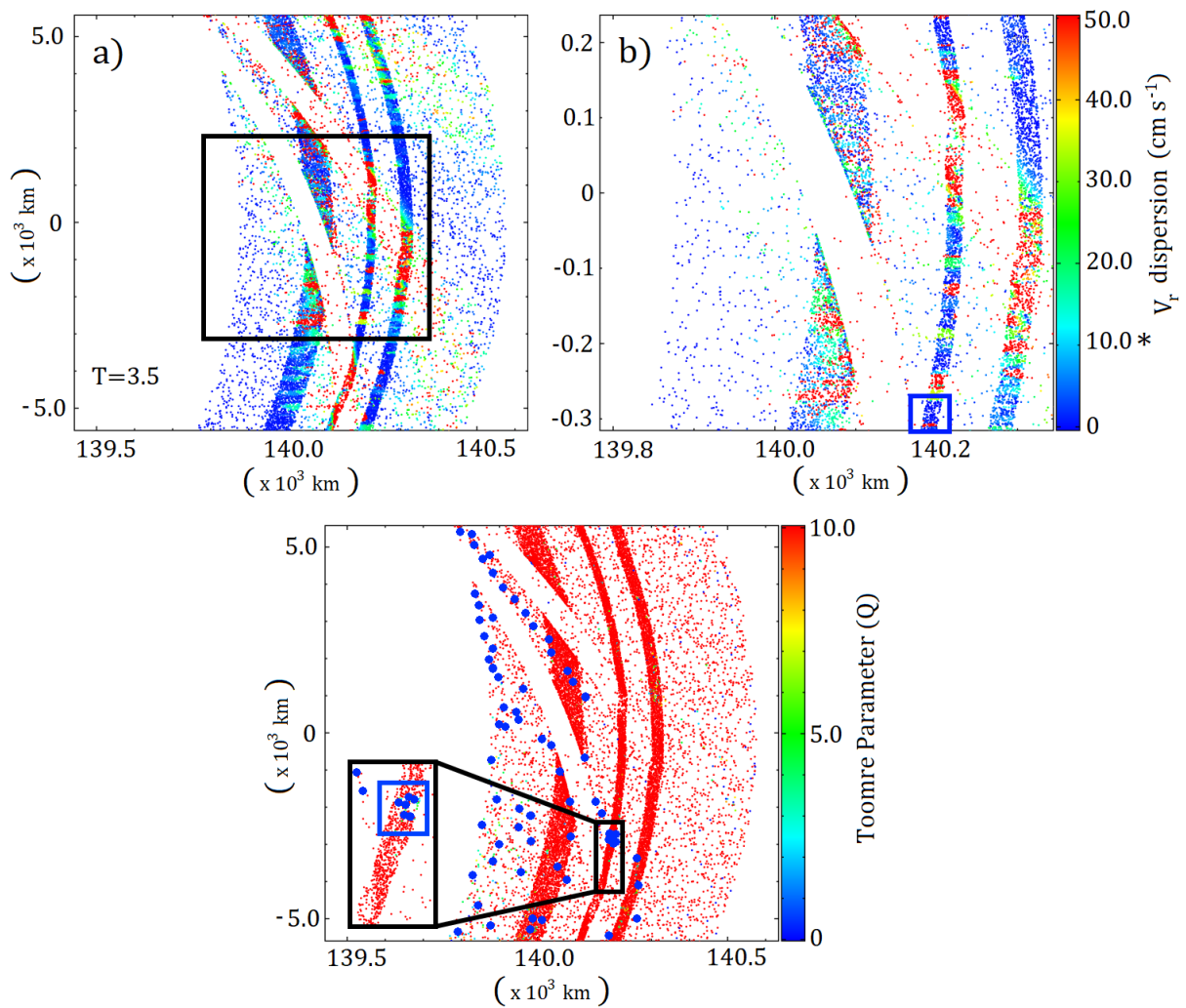

**Figure 6 |** a) Particles are plotted with their radial velocity dispersions dictating the colour for the model where the F ring core placed at its current location (~140,000 $km$). b) Zoomed section of (a)

shown by the black box. The star next to $10\ cms^{-1}$ represent the approximate escape velocity for an $150m$ radius icy moonlet. Thus, all particles coloured blue are below the escape velocities for the known population of $\sim 100m$ sized moonlets in the F ring. The highest radial velocity dispersions are again observed in or around the channel edges. However, the areas previously identified as having the maximum number densities are in locations below the escape velocity for an $150m$ icy moonlet (shown as a blue box). c) The Toomre parameter is calculated and plotted for all particles. It is clear to see that most of the ring is more than the stability criteria $Q < 1$ where the ring would be unstable to gravitational collapse. Nonetheless, small groups of particles do exist with $Q < 5$; the most significant of these is shown with the zoom section and highlighted with a blue box. Half of this group of particles satisfies $Q < 2$ and resides in the same location as the maximum number densities (Fig 2 b). Additionally, the particles that satisfy $Q < 2$ and that are associated with the oldest channel formed (channels formed from the first encounter since the start of the simulation) are enlarged on the plot.

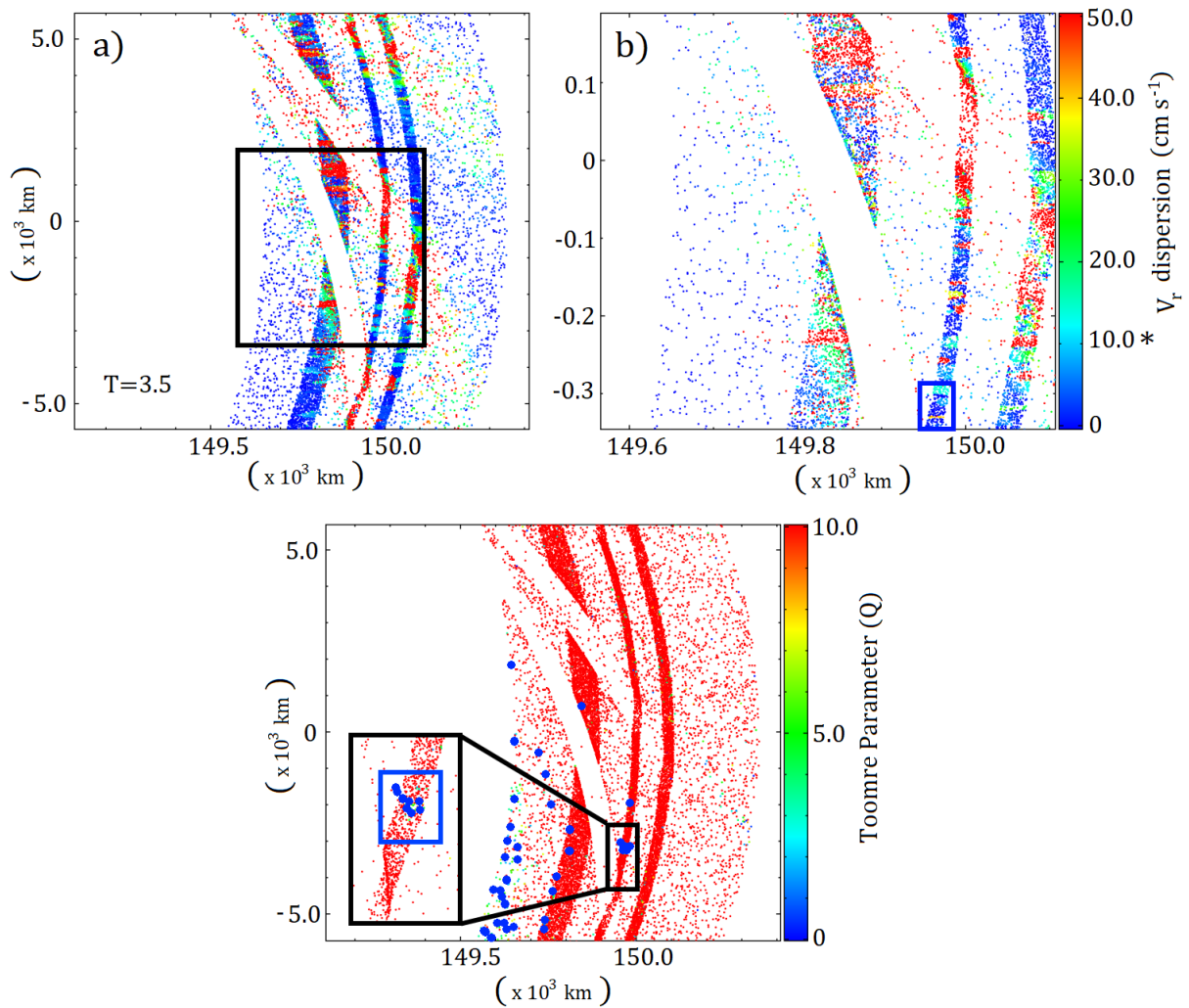

**Figure 7 |** a) Particles are plotted with their radial velocity dispersions dictating the colour for the model where the F ring core is set to $150,000 km$. b) Zoomed section of (a) shown by the black box. The star next to $10\ cms^{-1}$ represents the approximate escape velocity for a $150m$ radius icy moonlet. Thus, all particles coloured blue are below the escape velocities for the known population of $\sim 100m$ sized moonlets in the F ring. The highest radial velocity dispersions are again observed in or around the channel edges. The areas previously identified with the maximum number densities are in locations below the escape velocity for a $150m$ icy moonlet (shown as a blue box). c) The Toomre parameter is calculated and plotted for all particles. It is clear to see that most of the ring is above the stability criteria $Q < 1$ where the ring would be unstable to gravitational collapse. As with Fig 6c, two main groups of particles fall below $Q < 5$. The first is shown with the zoom section and coincides with the same location in Fig 3b as the maximum number densities, however, this time the groups of particles are larger. Additionally, the particles that satisfy $Q < 2$ and that are associated with the oldest channel formed (channels formed from the first encounter since the start of the simulation) are enlarged on the plot.

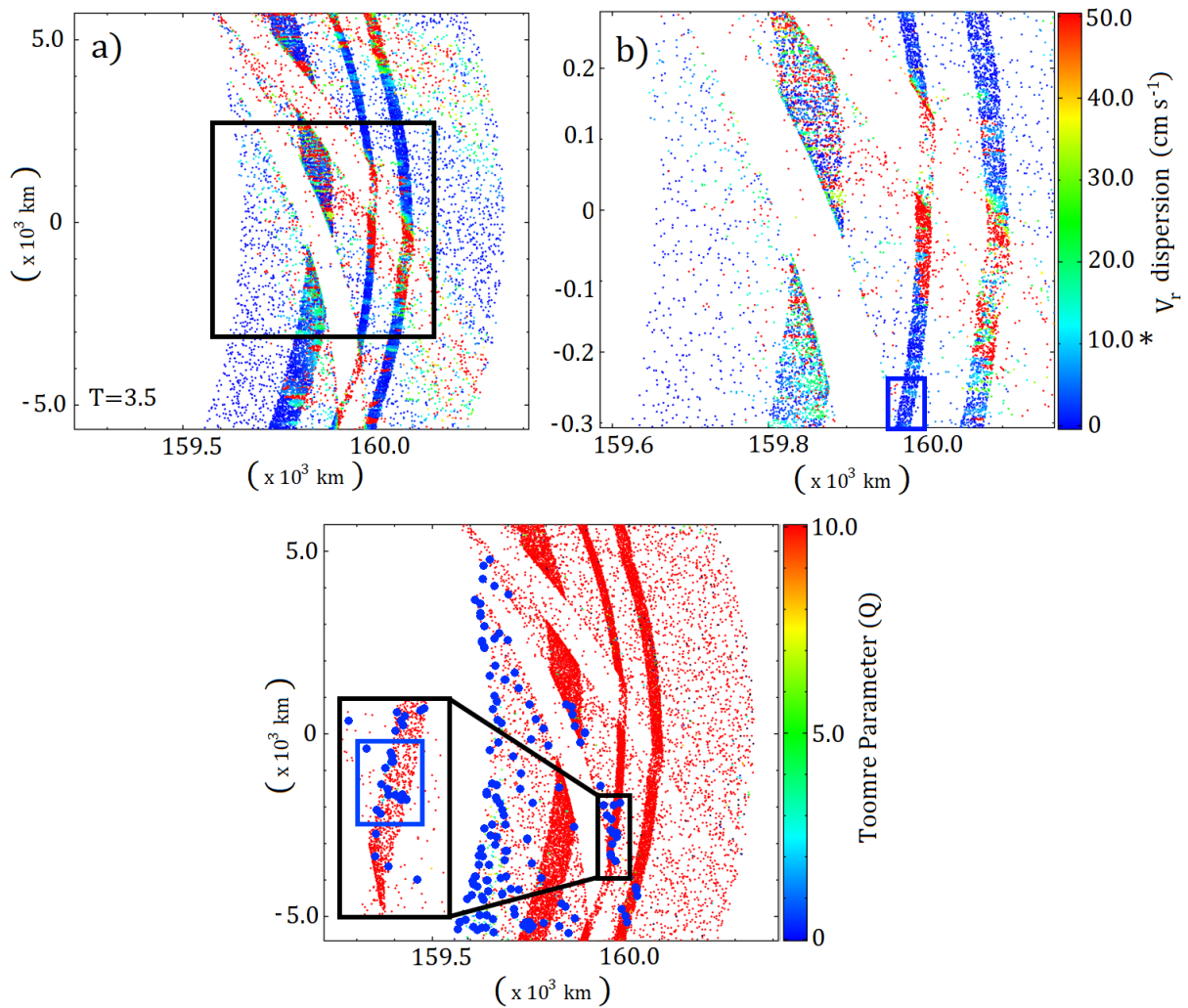

**Figure 8 |** a) Particles are plotted with their radial velocity dispersions dictating the colour for the model where the F ring core is set to $160,000 km$. b) Zoomed section of (a) shown by the black box. The star next to $10\ cms^{-1}$ represent the approximate escape velocity for a $150m$ radius icy moonlet. All particles coloured blue are below escape velocities for the known population of $\sim 100m$ sized moonlets in the F ring. The blue box shows the area where the maximum number densities are located but are also where radial velocity dispersions are below the escape velocity for a $150m$ icy moonlet. c) The Toomre parameter calculated and plotted for all particles is. It is easy to see that most of the ring is more than the stability criteria $Q < 1$ where the ring would be unstable to local gravitational collapse. As with Fig 6c & 7c two main groups of particles fall below $Q < 5$. The first is shown with the zoom section and coincides with the same location in Fig 4b as the maximum number densities. The groups of particles in this model cover the largest area and have the most members, suggesting that along with the highest maximum number densities of all models, accretion would be the most favourable when considering the disrupted area in the F ring. Additionally, the particles that

satisfy $Q < 2$ and that are associated with the oldest channel formed (channels formed from the first encounter since the start of the simulation) are enlarged on the plot.

From the snapshots (Fig 6, 7 & 8) we note some key findings. Firstly, as the F ring is moved outwards, more particles are seen exhibiting $Q < 2$. They are in the same groups within the streamer-channel formations, but the groups are larger. Secondly, there are two main areas where particles are seen with $Q < 2$ in all models; the first is where the maximum number densities are also witnessed, and the second area is to the bottom left of the F ring in Fig 6c, 7c & 8c. The latter is purely due to the fact this area is at the very edge of where Prometheus has distorted the ring and is thus closer to the case of a non-disrupted ring. Thirdly, the area where particles exhibit $Q < 2$ increases in size as the F ring moves radially outward. Coupled with the fact that higher maximum number densities occur in the same locations as the F ring moves radially outward, this would suggest accretion to be more favourable in these models. Finally, in all models the areas of the ring that do not encounter Prometheus have average Toomre values below $Q < 1$, however, due to the reduced orbital frequency at larger radial distances, the average $Q$ values will also decrease. This naturally increases the critical radial velocity dispersion particles can have before gravitational instabilities no longer occur.

## 5. DISSCUSSION

Radial velocities are important in the context of our work as it is the radial component in a particle's velocity which results in the highest encounter velocities with other particles, and Encounter velocities or impact velocities between particles are a key parameter in the aggregation of material in a planetary ring. The aggregation probability of colliding particles decreases significantly once impact velocities approach the escape velocities of the particles, and at velocities greater than the escape velocity, gravity becomes less important than the physical collision itself. Here the escape velocity for particles in a Saturnian ring (Goldreich et al 2004) is given as

$$v_{esc} = \sqrt{\frac{8\pi G\rho r^2}{3}} \qquad [6]$$

Although the F ring is known to have small $\mu m$ sized particles (Bosh et al 2002) the core is also populated with larger $\sim 100 m$ sized objects (Attree et al 2014, Attree et al 2012). Assuming this size distribution and an internal density of water ice ($\rho = 0.934\ gcm^{-3}$, which is consistent with the composition of Saturn's rings (Hedman et al 2013; Esposito 2010)), it would be expected that escape velocities in the F ring would be $v_{esc} \sim 1 \times 10^{-9} - 7.5 \times 10^{-2} ms^{-1}$.

The radial velocity dispersions generated by our models, from a direct result of the interaction of Prometheus on the F ring, exceed escape velocities of most F ring particles, should any collide. Despite this there are locations where radial velocity dispersions dip below the escape velocity for some $100 m$ sized moonlets; this occurs around the edges of the channels in all our models and is consistent with that reported by Beurle et al 2010. As the tidal environment decreases (F ring moves radially outwards), larger areas within the streamer-channel formations occur where particles fall below either the critical radial velocities dispersions defined by the Toomre parameter or escape velocities for larger particles. Coupled with the increasing maximum densities observed as the tidal environment decreases (Fig 1), this would drive further accretion of any imbedded moonlets at those locations.

In a ring not disrupted by Prometheus, we find that the Toomre parameter is $Q \ll 1$. However, it should be noted that no part of the real F ring will remain free from gravitational disruption. Due to the faster orbit of Prometheus, it will encounter the same part of the F ring every 110 orbits or 67.6 days. Ultimately, significant radial velocity dispersions will likely be introduced to all ring particles such that the average Toomre value for the F ring is $Q > 2$.

Conversely, as shown in our models, areas within the streamer-channel formations do indeed fall below $Q < 2$ and, as previously demonstrated, local condensations are likely to occur (Beurle et al 2010). Timescales for the growth of a 100m moonlet in the unperturbed F ring can be given as (Hyodo & Charnoz 2017)

$$T_{grow} \sim \frac{\rho R}{\Omega \Sigma (1+F_{grav)})} \qquad [7]$$

We assume a radius $R = 100m$, a surface density $\Sigma = 50\ kgm^{-2}, \rho = 934\ kg\ m^{-3}$ (density of ice) and $\Omega = 1.172 \times 10^{-4} s, 1.059 \times 10^{-4} s,\ 9.616 \times 10^{-5} s$ for the three models located at $140,000 km, 150,000\ \&\ 160,000 km$ respectively. Calculated timescales for growth of a $100m$ radius object in an unperturbed F ring is $T_{grow} \sim 0.25, 0.27, 0.31\ yr$ for the three models, located at $140,000 km, 150,000 km, 160,000 km$ respectively. Since the only variable we consider is the orbital frequency, the accretion time scales increase with increasing distance from Saturn. Comparing to locations in the streamer-channel formations that do become gravitationally unstable ($Q < 1$), we would expect that time scales would be larger than those calculated for the unperturbed F ring. This is purely because areas identified as having $Q < 1$ on the channel edges can also become gravitationally stable ($Q > 2$) during the natural cycle between streamer and channels over one orbital period of Prometheus. Here, radial velocity dispersions will remain at their minimum when channels are at their most open but increase as the channels fill back in with particles. The accretion rate in this scenario would be cyclic in the same way that the local density fluctuates. Additionally, Prometheus will encounter the same region of the F ring approximately 110 orbital periods later, so fragmentation could occur where accretion had been taking place. Since this will occur at timescales shorter than those calculated above, to accrete a $100m$ object ($0.17\ yrs$ when the F ring is radially located $\sim 140,000\ km$) we would expect timescales to be significantly larger for an F ring perturbed by Prometheus compared with an unperturbed ring.

Since our models assumed very low surface mass, it is worth noting some likely outcomes of models employing larger surface masses. It is expected that models with greater surface masses resist the natural cycle of maximum number densities observed in Fig [1] at the channel edges. More specifically, internal gravitational attraction within enhanced regions would act to reduce the expansion enhanced regions when Prometheus is at periapsis. As the surface mass increases the probability that Prometheus generates regions above the local Roche density, and local gravitational instabilities increases. The local Roche density of the F ring is given as $3\rho(R/a)^3 = 0.15\ gcm^{-3}$, where $a$ is the semi-major axis, $R$ is the radius of the planet and $\rho$ is the density of the planet (Beurle et al 2010). Also, according to Eq [7], time scales of 100m moonlet growth would occur before Prometheus encounters the same segment of the F ring ($0.17\ yrs$). This occurs as the F ring approaches surface densities more comparable to that of the main A and B rings, $\Sigma > 100\ kgm^{-2}$. The growth of 100m sized moonlets in the F ring in time frames shorter than a second encounter of Prometheus would have a significant effect on the moonlet population.

The work presented in this manuscript assumed an F ring with a core and inner and outer strands with initial conditions comparable to Murray et al. 2005. The model employed in Beurle et al. 2010 only considered a 40km radially wide ring with no additional features such as strands or core. We note some differences in the early evolution post encounter that would not have been seen by Beurle et al. 2010. A surge in maximum number densities was observed to occur after 5 orbital periods post encounter in Beurle et al. 2010. This surge is thought to be caused by Keplerian shear on the streamer-channel formations which causes overlapping of particle trajectories around the channel edges. The Keplerian shearing decreases spacing between each streamer-channel and particles are seen to bunch closer at the channel edges as the ring evolves.

When considering the F ring model used by Murray et al 2005 (the same as ours located at ~140,000km), the core is not radially disrupted to the point that it overlaps the outer strand. However, as the F ring moves radially outwards (150,000km & 160,000km), the core is radially disrupted such that there is an overlapping of the outer strand and core. This happens very early post encounter and before the surge reported by Beurle et al. 2010. Therefore, in the context of an outwardly migrated F ring, it is important to consider the early evolution of a stranded F ring and the crucial role played in the evolution of the local particle density by Prometheus.

In conclusion, we find that in a scenario where the F ring migrates radially outwards, the streamer-channel formations that are responsible for local condensations (Beurle et al 2010) might experience a further enhancement. Maximum number densities increase at channel edges as the tidal environment decreases (Fig 1). This also coincides with the same locations where the local Toomre parameter of the ring becomes unstable to gravitational collapse, or radial velocity dispersions are below the escape velocities of larger moonlet sized particles. As the tidal environment decreases, larger areas of the ring are seen to fall below the critical values for radial velocity dispersions where the maximum number densities are observed. The result is that both the increase of maximum number densities and an increased fraction of particles with $Q < 2$ within a decreasing tidal environment can further enhance local condensations.

## 6. REFERENCE

Attree N. O. et al. 2014, Icarus, 227, 56 – 66

Attree, N. O., Murray, C. D., Cooper, N. J. & Williams, 2012, APJL, 755.

Beurle, K. et al., 2010, ApJL, 718, 176 – 180.

Bosh, A. S., et al. 2002, Icarus, 157, 1, 57 – 75.

Chandrasekhar, S. 1969, Yale University Press, New Haven.

Charnoz, S., Crida, A., Castillo-Rogez, J. C., Lainey, V., Dones, L., Karatekin, Ö., Salmon, J. 2011, Icarus, 216, 2, 535 – 550.

Charnoz, S., Salmon, J., Crida, A. 2010, Nature, 465, 7299, 752 – 754.

Cuzzi, J.N., Whizin, A.D., Hogan, R.C., Dobrovolskis, A.R., Dones, L., Showalter, M.R., Colwell, J.E. and Scargle, J.D., 2014. Icarus, 232, pp.157 – 175.

Dubinski, J., 2017, arXiv preprint arXiv:1709.08768.

Esposito, L.W., 2010, Annual Review of Earth and Planetary Sciences, 38, pp.383-410.

Goldreich, P. et al., 2009, Annu. Rev. Astron. Astrophys. 42, 549 – 601.

Hedman, M. M., Burns, J. A., Hamilton, D. P., Showalter, M. R., 2013, Icarus, 223, 252 – 276.

Hedman, M.M. et al. 2011, Icarus, 215, 695 – 711.

Hedman, M.M. et al., 2013, Icarus, 223(1), pp.105-130.

Hedman, M.M., Nicholson, P.D., Salo, H., Wallis, B.D., Buratti, B.J., Baines, K.H., Brown, R.H., Clark, R.N., 2007, AJ, 133, 2624 – 2629.

Hedman, Matthew M., et al. 2007, Science, 317, 5838, 653 – 656.

Horn, L.J., Showalter, M.R., Russell, C.T., 1996, Icarus, 124(2), pp.663-676.

Hyodo, R, Charnoz, S., 2017. arXiv preprint arXiv:1705.07554.

Hyodo, R., Ohtsuki, K. 2014, ApJ, 787, 56.

Hyodo, R., Ohtsuki, K. 2015, Nature Geoscience, 8, 686 – 689.

Lewis, M.C., Stewart, G.R., 2009, Icarus, 199, 387 – 412.

Lewis, M.C., Stewart, G.R., 2009. Icarus, 199(2), pp.387-412.

Murray, C. D., et al. 2005, Nature, 437, 7063, 1326.

Murray, C. D., et al. 2008, Nature, 453, 7196, 739.

Price D.J., 2007, PASA, 24, 159.

Rein H., Latter H. N., 2013, MNRAS, 431, 1, 145 – 158.

Rein H., Papaloizou J. C. B., 2010, A&A, 524, A22.

Salo H., Schmidt J., Spahn F., 2001, Icarus, 153, 2 295 – 315.

Salo, H., 1992, Nature, 359(6396), p.619.

Seiß, M. et al., 2017, arXiv preprint arXiv:1701.04641.

Springel V., 2005, MNRAS, 364, 1105.

Sremčević, M. et al., 2007, Nature, 449(7165), p.1019.

Sutton P. J., Kusmartsev F. V, 2013, Sci Rep. 3, 1276.

Sutton P. J., Kusmartsev F. V, 2014, MNRAS, 439, 1313 – 1325.

Sutton P. J., Kusmartsev F. V, 2016, Earth, Moon, and Planets. 1 – 15.

Thomas, P.C. 2010, Icarus, 208(1), pp.395-401.

Thomson et al., 2007, Gephys. Res. Lett. 34, 24, L24203.

Tiscareno M. S. et al., 2010, ApJ, 718, L92 – L96.

Tiscareno, M.S. et al., 2008, 135(3), p.1083.

Toomre, A., 1964, ApJ, 139, 1217 – 1238.

Weiss, J.W., Porco, C.C., Tiscareno, M.S., 2009, AJ, 138, 1, 272 – 286.

**Author contributions**

P. J. S performed the numerical simulations, analysed the results and wrote the manuscript.

**Competing financial interests**

The author declares no competing financial interests.

**Acknowledgements**

The author would like to thank the anonymous referee for their useful comments which greatly improved the quality of the manuscript.